# Investigation of flow field characteristics and performance of carbon-hydrogen/oxygen-rich air rotating detonation engine

Guangyao Rong(荣光耀), Miao Cheng(程杪), Yunzhen Zhang(张允祯), Zhaohua Sheng(盛兆华), Jianping Wang(王健平)*

*Center for Combustion and Propulsion, CAPT & SKLTCS, Department of Mechanics and Engineering Sciences, College of Engineering, Peking University, Beijing 100871, China*

**Abstract:** Numerical simulations were conducted to investigate the flow field characteristics and performance of a carbon-hydrogen/oxygen-rich air rotating detonation engine (RDE). Three distinct flow field structures were observed in the gas-solid two-phase RDE. The results show that reducing the hydrogen equivalence ratio and particle diameter both contribute to the transition from gas-phase single-front detonation to gas-solid two-phase double-front detonation and further to gas-solid two-phase single-front detonation. The effects of solid fuel particle diameter and hydrogen equivalence ratio on the flow field characteristics and performance are revealed. The results show that reducing the particle diameter enhances the speed of the two-phase detonation wave, improves the pressure gain in the combustion chamber, and increases the specific impulse. Decreasing the hydrogen equivalence ratio reduces the detonation wave speed, enhances the stability of the detonation flow field, increases the pressure gain in the detonation wave and combustion chamber and boosts thrust. Furthermore, the selection of operational conditions to ensure stable operation and optimal performance of the RDE is discussed. In order to take into account the requirements of stability, pressure gain performance and propulsion performance, two-phase single-front detonation should be realized in gas-solid two-phase RDE, and smaller hydrogen equivalent ratio and appropriate particle diameter should be selected. According to the conclusion of this study, the particle diameter should be 0.5-1 μm. Under such conditions, the detonation flow field demonstrates good stability, allowing the RDE to achieve higher pressure gain and specific impulse while maintaining stable operation.

*Corresponding author
Email address: **wangjp@pku.edu.cn** (Jianping Wang(王健平))

# 1. Introduction

The detonation wave is a self-sustaining supersonic combustion wave formed by the close coupling of the leading shock wave and the chemical reaction zone. Detonation combustion exhibits a thermal efficiency of 13% higher than that of isobaric combustion based on Brayton cycle [1]. The utilization of detonation in aerospace engines holds the potential to break through the bottleneck of current propulsion system performance. Among the various types of aerospace engines, the rotating detonation engine (RDE) emerges as a highly promising option, employing detonation as the combustion mode. The RDE demonstrates several advantages, including small entropy increase, high thermal efficiency, simplified structure, and only a single ignition during startup.

In the 1960s, Voitsekhovskii first proposed the basic concept of rotating detonation waves, and successfully obtained six rotating detonation waves in a disk combustor [2]. Since the 1980s, Bykovskii et al. have conducted a series of studies on RDE, including the configuration, fuel, injection mode, flow field structure, working mode and performance of RDE [3–7]. Currently, numerous international research institutions and researchers have carried out experimental research, numerical simulation and theoretical research on RDE, including visualization [8–17], modal and stability [18–23], multiphase detonation [24–30], nozzle and performance [31–34], injection mode [35], combustor configuration [36–41], etc. Owing to its vast prospects and inherent advantages, RDE has become an international research hotspot.

Extensive research has been conducted on gas-phase RDEs, including hydrogen-air [42], hydrocarbon fuel-air [43], gaseous kerosene-air [44], etc. There was a recent work that discusses liquid-detonation interactions in RDEs, including ZND model and liquid dispersion measurements. Kumar et al. used ozone and hydrogen peroxide as trace ignition promoters to perform the ZND computations of RP2-1-O2 and RP2-2-O2 propellants [45]. Zhao et al. studied the influence of the outlet convergence ratio of the combustion chamber on the rotating detonation wave of liquid kerosene [46]. Meng et al. studied the propagation mode of liquid kerosene/oxygen-enriched air rotating detonation wave [47]. Hoeper et al. used the method of planar laser-induced fluorescence to optically measure the refilling kinetics of kerosene in RDE [48]. However, research on gas-solid two-phase RDE is still relatively rare. The simulation of two-phase RDE is challenging because of the need to consider the coupling of fluid dynamics and chemical reactions as well as the coupling of gas and solid phases, which leads to the complexity of the two-phase rotating detonation flow field. Compared with gaseous fuels, solid fuels are more difficult to mix with oxidants and more stable, which makes it challenging to form gas-solid two-phase rotating detonation waves at room temperature. However, adding solid fuels such as carbon particles to gas-phase RDE has exciting advantages: it can reduce the cost of propellant; the volumetric calorific value of solid fuel is high;

solid fuel is safe and easy to store. Gas-solid two-phase RDE can be used in aerospace engines, such as rocket engines, ramjet engines, etc. Given gas-solid two-phase RDE has the above remarkable advantages, it is very meaningful to study gas-solid two-phase RDE.

As a low-cost and stable solid fuel, coal can be used in gas-solid two-phase RDE. Bykovskii et al. first experimentally obtained a continuous rotating detonation wave with hydrogen-coal-air as propellant. The diameter of the solid particles is 1-7 μm, and the detonation wave velocity is 1.86-1.1 km/s [49]. They then continued to do a series of experiments and found that volatile components and additives in coal had a significant effect on detonation waves [50,51]. Dunn et al. conducted experiments on Carbon Black-hydrogen-air RDE, and found that the addition of Carbon Black can increase the detonation wave velocity, and the mass of carbon particles has a linear relationship with the combustion heat [52]. They then confirmed through further experiments that the addition of carbon particles can broaden the range of operating parameters for detonation [53]. Numerical simulations were also performed to study gas-solid two-phase RDE. Salvadori et al. studied the effect of burning coal particles on RDE by Euler-Euler method, and found that the addition of particles did not change the flow field significantly, and the coupling effect between the two phases needed to be further studied [54]. Zhu et al. studied the effects of hydrogen flow rate and solid particle diameter on coal-hydrogen-air RDE, and found that the rotating detonation wave mainly relied on the reaction of hydrogen-air, and coal particles could not form detonation [28]. They then numerically simulated the carbon-air RDE and studied the influence of the equivalence ratio on the flow field [29]. It was found that a low-temperature air gap would appear in the flow field, and a higher equivalence ratio would cause two detonation waves.

So far, research on gas-solid two-phase rotating detonation engines remains limited, and there are many issues that require urgent investigation. The fluid physics of gas-solid two-phase rotating detonation waves are still not well understood. Specifically, there is a lack of research on the flow field structure, dynamics of detonation mode transition, instability, pressure gain performance, and propulsion performance of gas-solid two-phase RDEs, both nationally and internationally. Mechanisms and practical applications of gas-solid two-phase RDEs require a clear understanding of the influence of different factors on the flow field structure, dynamics of detonation mode transition, and instability. It is crucial to balance the engine's operational stability and performance output, so that the gas-solid two-phase RDE can achieve a good operation state. Therefore, exploring these issues holds significant importance for understanding the operational mechanism, working characteristics, and practical applications of gas-solid two-phase RDEs.

In summary, the structure of this paper is as follows. Firstly, the flow field characteristics, wave structure,

and instability of gas-solid two-phase rotating detonation engines were investigated. Secondly, the effects of solid fuel particle diameter and hydrogen equivalence ratio on the self-sustained propagation, mode transition, propagation characteristics, and stability of two-phase rotating detonation waves were revealed. Thirdly, the pressure gain characteristics of the detonation wave and combustion chamber and propulsion performance were studied concerning changes in particle diameter and hydrogen equivalence ratio. Furthermore, the selection of operational conditions to ensure stable operation and optimal performance of the RDE was discussed.

## 2. Computational Methods

In this study, the Eulerian-Lagrangian method is used to simulate the two-phase rotating detonation combustion. Detonation is a transient phenomenon involving multi-species chemical reactions. For the gas phase, the compressible Navier-Stokes equations are used as the governing equations, including the mass conservation equation of each species and source terms. Since the dilute particles are considered in this study, the volume fraction effects of the discrete phase are ignored for the gas phase control equation [25,55]. The governing equations are as follows:

$$\frac{\partial \rho}{\partial t} + \nabla \cdot (\rho \boldsymbol{U}) = S_{mass} \tag{1}$$

$$\frac{\partial \rho \boldsymbol{U}}{\partial t} + \nabla \cdot (\rho \boldsymbol{U}\boldsymbol{U}) + \nabla p = \nabla \cdot \boldsymbol{\tau} + \boldsymbol{S}_{mom} \tag{2}$$

$$\frac{\partial \rho E}{\partial t} + \nabla \cdot [(\rho E + p)\boldsymbol{U}] = \nabla \cdot (\lambda \nabla T) + \nabla \cdot (\boldsymbol{\tau} \cdot \boldsymbol{U}) + \dot{\omega}_T + S_{energy} \tag{3}$$

$$\frac{\partial \rho Y_k}{\partial t} + \nabla \cdot [\rho \boldsymbol{U} Y_k] + \nabla \cdot \boldsymbol{s}_k = \dot{\omega}_k + S_{species,k} \tag{4}$$

And the ideal gas state equation, as shown below:

$$p = \rho R T \tag{5}$$

Where $t$ denotes time; $\rho$ is the fluid density; $\boldsymbol{U}$ is velocity vector; $p$ is the fluid pressure; $E$ is total energy; $\lambda$ is the thermal conductivity of the fluid; $T$ is the fluid temperature; $R$ is gas constant; $Y_k$ is the mass fraction of $k$th species; $s_k$ is the mass diffusion rate of $k$th species; $S_{mass}$, $\boldsymbol{S}_{mom}$, $S_{energy}$ and $S_{species,\ k}$ denote mass, momentum, energy and species interphase exchanges, respectively. $\boldsymbol{\tau}$ is a viscous stress tensor, defined as

$$\boldsymbol{\tau} = \mu[\nabla \boldsymbol{U} + (\nabla \boldsymbol{U})^T - \frac{2}{3}(\nabla \cdot \boldsymbol{U})\boldsymbol{I}] \tag{6}$$

$\dot{\omega}_k$ is the reaction rate of *k*th species; $\dot{\omega}_T$ is the heat released by combustion, defined as

$$\dot{\omega}_T = -\sum_{k=1}^{N} \Delta h_{f,k}^0 \dot{\omega}_k \tag{7}$$

Where $\Delta h_{f,k}^0$ is the formation enthalpy of *k*th species.

For the solid phase, the Lagrangian method is used to track the discrete solid phase composed of a large number of spherical solid particles [56]. In this study, the Eulerian-Lagrangian method was employed using the point-force approximation, which means that the diameter of the solid phase Lagrangian particles should be smaller than the size of Eulerian cells [55]. Under this criterion, the gas phase physical quantities near the solid particles can be wall calculated by the interpolation of the solid particles position in the Euler grid [57]. The calculated gas phase quantities are crucial for capturing the two-phase coupling effects. Sontheimer et al. [58] and Luo et al. [59] suggested that the ratio between the size of Eulerian cells and the size of Lagrangian particles should be greater than 10. In this study, this ratio ranges from 17 to 200, which is close to or significantly greater than 10. Therefore, the selected size of the Eulerian grid in this study effectively simulates the gas-solid two-phase coupling and the dynamic characteristics of solid particles, thus enabling an effective simulation of the rotating detonation flow field.

Due to the consideration of dilute particles (volume fraction<0.001) in this study, the interactions between particles are neglected for solid phase [25,55]. It is worth stating that this condition is applicable in the combustion chamber, however, for non-premixed systems, this may be totally untrue near the injector. In this study, near the head of the combustion chamber, the solid particle volume fraction reaches its maximum value of 0.000623. This value falls within the range considered as dilute particles. Therefore, dilute particles can be considered. The fuel was introduced using the full-area intake method, without considering the injection mixing structure of the RDE. The state of the particles after they have been diluted is considered, so the point of dilute particles is applicable. Therefore, when employing the full-area intake method, dilute particles can be considered, but when considering the injection mixing structure of RDE, the presence of dense particles must be considered. The governing equations of mass, momentum and energy of a single solid particle in the discrete solid phase are as follows:

$$\frac{dm_p}{dt} = \dot{m}_p \tag{8}$$

$$\frac{d\boldsymbol{u}_p}{dt} = \frac{\boldsymbol{F}_p}{m_p} \tag{9}$$

$$C_{p,p} \frac{dT_p}{dt} = \frac{\dot{Q}_p + \dot{Q}_h}{m_p} \tag{10}$$

Where $t$ is the time, $m_p$ is the mass of the solid particle, $\dot{m}_p$ is the mass change of the solid particle, $u_p$ is the velocity of the solid particle, $F_p$ is the drag force of the solid particle, $C_{p,p}$ is the specific heat capacity of the solid particle, $\dot{Q}_p$ is the convective heat transfer between the gas phase and the solid phase, $\dot{Q}_h$ is the combustion heat release for heating the particles. The surface reaction model is used for carbon combustion [60,61], and the expression of $\dot{m}_p$ is as follows:

$$\dot{m}_p = -A_p P_{ox} \frac{D_0 R_k}{D_0 + R_k} \tag{11}$$

$A_p$ represents the surface area of the solid particles, and $P_{ox}$ represents the oxidant partial pressure of the gas around the solid particles. $D_0$ and $R_k$ are the diffusion rate coefficient and kinetic rate, respectively. The expressions are as follows:

$$D_0 = \frac{C_1}{d_p} (\frac{T_p + T_\infty}{2})^{0.75} \tag{12}$$

$$R_k = C_2 e^{-\frac{E_p}{RT_p}} \tag{13}$$

The drag force $F_p$ of solid particles is calculated by the following equation:

$$F_p = m_p \frac{u - u_p}{\tau_r} \tag{14}$$

$u$ is the gas velocity. $\tau_r$ is the relaxation time of particles [62], calculated by the following equation:

$$\tau_r = \frac{\rho_p d_p^2}{18\mu} \frac{24}{C_d \text{Re}_d} \tag{15}$$

$\rho_p$ is the particle density and $d_p$ is the particle diameter. $C_d$ is the drag coefficient, $Re_d$ is the relative Reynolds number, calculated by the following equation:

$$C_d = \begin{cases} 0.424, & \text{Re}_d > 1000 \\ \frac{24}{\text{Re}_d}(1 + \frac{1}{6}\text{Re}^{\frac{2}{3}}), & \text{Re}_d \leq 1000 \end{cases} \tag{16}$$

$$\text{Re}_d = \frac{\rho d_p |u_p - u|}{\mu} \tag{17}$$

Convective heat transfer $\dot{Q}_p$ is calculated as follows:

$$\dot{Q}_p = h_c A_p (T - T_p) \tag{18}$$

$T_p$ is the temperature of solid particles. $h_c$ is the convective heat transfer coefficient, calculated according to Nusselt number:

$$Nu = \frac{h_c d_p}{k_e} = 2.0 + 0.6 \operatorname{Re}_d^{\frac{1}{2}} \operatorname{Pr}^{\frac{1}{3}} \tag{19}$$

$k_e$ is the thermal conductivity of fluid, and $Pr$ is the gas Prandtl number.

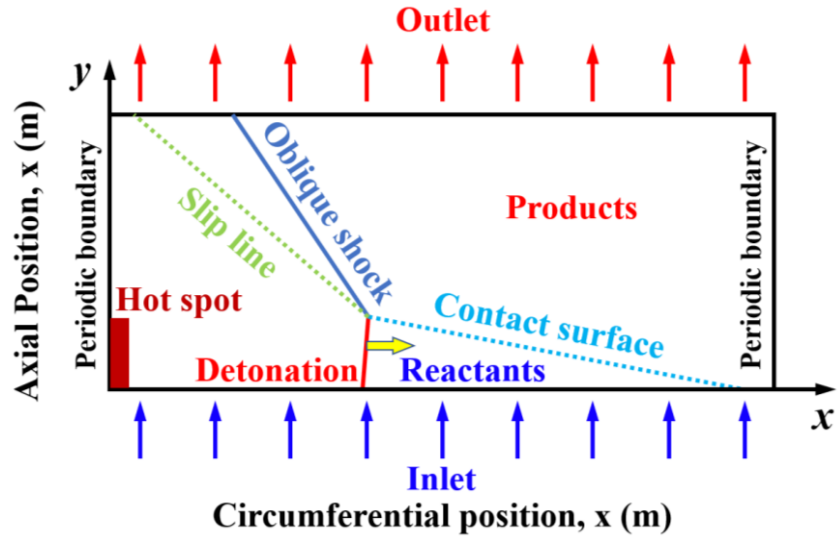

**Fig. 1 The computational domain of two-dimensional rotating detonation engine.**

In this study, the two-dimensional rotating detonation flow field with carbon-hydrogen as fuel is numerically simulated. When we study the configuration of the common coaxial annular cavity RDE, because the radial thickness of the RDE is much smaller than the diameter and axial length of the RDE, the flow field can be regarded as a two-dimensional configuration without thickness. This simplification has been widely used in the numerical simulation of RDE, including gas-phase RDE [63,64], gas-liquid two-phase RDE [26,30], gas-solid two-phase RDE [28,29], etc. Un-wrapped RDE simulations are an ab-initio step to understand condensed phase detonation.

Fig. 1 is a schematic diagram of the computational domain of a two-dimensional rotating detonation engine. In the expanded RDE calculation domain, the length of the inlet and outlet is 0.12 m, that is, the radius of the RDE is 0.038 m. The left and right sides of the computational domain are periodic boundary conditions, and the boundary length is 0.04 m, that is, the axial length of the RDE is 0.04 m. The gaseous fuel of RDE is hydrogen, the solid fuel is carbon particles, and the oxidant is oxygen-rich air. For gas phase hydrogen/oxygen-rich air, the inlet adopts full-area intake condition [64–66] for gas phase fuel injection. The injection total temperature of the gas phase is 360 K and the injection total pressure is 8 atm. For solid phase, the velocity of carbon particles injected into RDE is 50 m/s, the equivalence ratio of carbon particles is 2.29, and the diameter of carbon particles varies with different cases and working conditions. In general, the velocity of carbon particles injected into the RDE is not higher than that of the injected gas because the gas phase is more mobile than the solid phase, so the carbon particle injection velocity chosen for this study satisfies this condition. The mass fraction of oxygen in oxygen-rich air is 0.457. The chemical reaction mechanisms of hydrogen and carbon particles adopt detailed chemical reaction mechanism [67] and surface chemical reaction mechanism [29,68] respectively. The initiation method of RDE is to set hot spots at the initial time, as shown in Fig. 1. The pressure of the hot spot is 15 atm and the temperature is 2500 K. At the initial moment, the front of the hot spot is the reactant, the pressure is 1 atm, and the temperature is 300 K.

Table 1 Operating condition table (The oxidizer is Oxygen-rich air)

| Case number | Equivalence ratio (H2) | Equivalence ratio (carbon) | Solid particle diameter (μm) | Injection total pressure (atm) | Injection total temperature (K) | Whether successfully detonated |
| --- | --- | --- | --- | --- | --- | --- |
| 1 | 0.5 | 2.29 | 0.2 | 8 | 360 | Yes |
| 2 | 0.5 | 2.29 | 0.5 | 8 | 360 | Yes |
| 3 | 0.5 | 2.29 | 1 | 8 | 360 | Yes |
| 4 | 0.5 | 2.29 | 2 | 8 | 360 | Yes |
| 5 | 0.5 | 2.29 | 4 | 8 | 360 | Yes |
| 6 | 0.5 | 2.29 | 8 | 8 | 360 | Yes |
| 7 | 0.4 | 2.29 | 0.2 | 8 | 360 | Yes |
| 8 | 0.4 | 2.29 | 0.5 | 8 | 360 | Yes |
| 9 | 0.4 | 2.29 | 1 | 8 | 360 | Yes |
| 10 | 0.4 | 2.29 | 2 | 8 | 360 | Yes |
| 11 | 0.4 | 2.29 | 4 | 8 | 360 | Yes |
| 12 | 0.4 | 2.29 | 8 | 8 | 360 | Yes |
| 13 | 0.3 | 2.29 | 0.2 | 8 | 360 | Yes |
| 14 | 0.3 | 2.29 | 0.5 | 8 | 360 | Yes |
| 15 | 0.3 | 2.29 | 1 | 8 | 360 | Yes |
| 16 | 0.3 | 2.29 | 2 | 8 | 360 | Yes |
| 17 | 0.3 | 2.29 | 4 | 8 | 360 | Yes |
| 18 | 0.3 | 2.29 | 8 | 8 | 360 | Yes |
| 19 | 0.2 | 2.29 | 0.2 | 8 | 360 | Yes |
| 20 | 0.2 | 2.29 | 0.5 | 8 | 360 | Yes |
| 21 | 0.2 | 2.29 | 1 | 8 | 360 | Yes |
| 22 | 0.2 | 2.29 | 2 | 8 | 360 | Yes |
| 23 | 0.2 | 2.29 | 4 | 8 | 360 | No |
| 24 | 0.2 | 2.29 | 8 | 8 | 360 | No |
| 25 | 0.1 | 2.29 | 0.2 | 8 | 360 | Yes |

| Case number | Equivalence ratio (H2) | Equivalence ratio (carbon) | Solid particle diameter (μm) | Injection total pressure (atm) | Injection total temperature (K) | Whether successfully detonated |
|---|---|---|---|---|---|---|
| 26 | 0.1 | 2.29 | 0.5 | 8 | 360 | Yes |
| 27 | 0.1 | 2.29 | 1 | 8 | 360 | Yes |
| 28 | 0.1 | 2.29 | 2 | 8 | 360 | No |
| 29 | 0.1 | 2.29 | 4 | 8 | 360 | No |
| 30 | 0.1 | 2.29 | 8 | 8 | 360 | No |
| 31 | 0.3 | 2.29 | 0.7 | 8 | 360 | Yes |
| 32 | 0.3 | 2.29 | 1.3 | 8 | 360 | Yes |
| 33 | 0.3 | 2.29 | 1.6 | 8 | 360 | Yes |
| 34 | 0.3 | 2.29 | 2.4 | 8 | 360 | Yes |
| 35 | 0.3 | 2.29 | 2.8 | 8 | 360 | Yes |
| 36 | 0.3 | 2.29 | 3.2 | 8 | 360 | Yes |
| 37 | 0.3 | 2.29 | 3.6 | 8 | 360 | Yes |
| 38 | 0.3 | 2.29 | 4.4 | 8 | 360 | Yes |

Table 2 Operating condition table (The oxidizer is air)

| Case number | Equivalence ratio (H2) | Equivalence ratio (carbon) | Solid particle diameter (μm) | Injection total pressure (atm) | Injection total temperature (K) | Whether successfully detonated |
|---|---|---|---|---|---|---|
| 39 | 0.5 | 2.29 | 1 | 8 | 360 | No |
| 40 | 0.58 | 2.89 | 0.65 | 12 | 500 | Yes |
| 41 | 0.58 | 2.89 | 0.7 | 12 | 500 | Yes |
| 42 | 0.58 | 2.89 | 0.75 | 12 | 500 | Yes |
| 43 | 0.58 | 2.89 | 0.8 | 12 | 500 | Yes |
| 44 | 0.58 | 2.89 | 0.85 | 12 | 500 | Yes |
| 45 | 0.58 | 2.89 | 0.9 | 12 | 500 | Yes |
| 46 | 0.58 | 2.89 | 1 | 12 | 500 | No |
| 47 | 0.73 | 2.89 | 0.8 | 12 | 500 | Yes |
| 48 | 0.65 | 2.89 | 0.8 | 12 | 500 | Yes |
| 49 | 0.5 | 2.89 | 0.8 | 12 | 500 | No |

To study the influence of particle diameter and hydrogen equivalence ratio on the self-sustained propagation, wave structure, stability, pressure gain characteristics, and propulsion performance of the gas-solid two-phase rotating detonation wave, a series of cases numbered from 1 to 30 were conducted. The range of solid particle diameters was from 0.2 μm to 8 μm, and the range of hydrogen equivalence ratio was from 0.1 to 0.5. For quantitative analysis of the transition between rotating detonation modes, additional cases numbered from 31 to 38 were conducted. The range of solid particle diameters was from 0.7 μm to 4.4 μm, and the hydrogen equivalence ratio was set to 0.3. Table 1 presents the operational conditions using oxygen-enriched air as the oxidizer. Cases 39 to 49 involve air as the oxidizer, cases 40 to 46 vary the particle diameter, and cases 47 to 49 vary the hydrogen equivalence ratio. Table 2 provides the operational conditions using air as the oxidizer.

The actual studies that have demonstrated the feasibility of gas-solid two-phase RDE are the experimental research conducted by Bykovskii et al. [49] and Dunn et al. [53]. Bykovskii et al. obtained a detonation wave velocity range of 1100-1860 m/s in their experiments on gas-solid two-phase RDE, with air as the oxidizer and a carbon equivalence ratio ranging from 0.25 to 3.5 [49]. Dunn et al. obtained a rotating detonation wave velocity range of 1200-1700 m/s in their experiments on gas-solid two-phase RDE, with air as the oxidizer and a global equivalence ratio of 1 [53]. In this study, when the oxidizer is air and the carbon equivalence ratio is 2.89, the detonation wave velocity range is 1532-1655 m/s. The detonation velocities obtained through simulation fall within the experimental range and qualitatively agree with the experimental results.

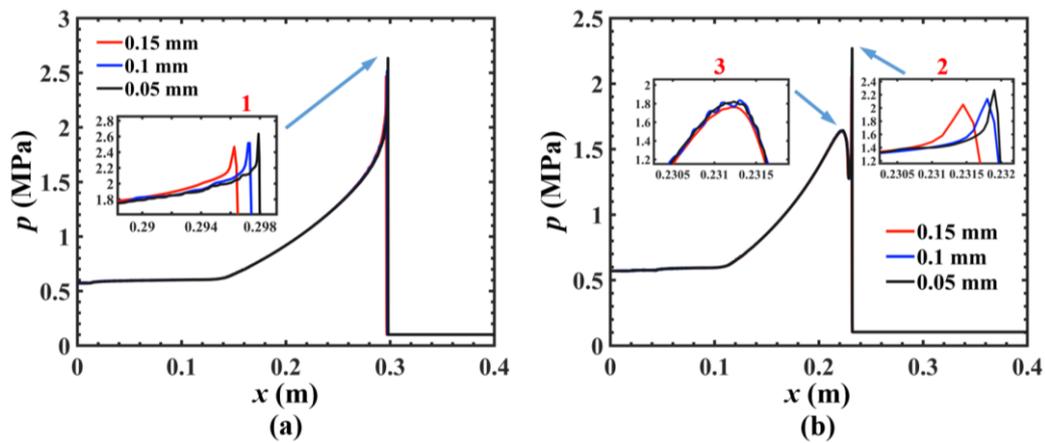

**Fig. 2 Pressure distribution of one-dimensional detonation tube. A total of two cases, each case uses three grid sizes: 0.05 mm, 0.1 mm, 0.15 mm. (a) Gas-solid two-phase single-front detonation. (b) Gas-solid two-phase double-front detonation.**

This study is conducted on the open source computational fluid dynamics platform OpenFOAM, using the in-house solver BYRFoam. The solver has been introduced and verified in previous studies [64,66,69]. The 1D detonation tube and 2D RDE flow field are simulated to verify the program and grid independence. The case in Fig. 2(a) corresponds to the following conditions: hydrogen equivalence ratio is 0.7, carbon particle equivalence ratio is 0.32, carbon particle radius is 0.1 μm, oxygen mass fraction in oxygen-rich air is 0.283, initial pressure of reactant is 1 atm, initial temperature is 300 K. As shown in the figure, the detonation wave in this case has only one pressure peak, that is, the gas phase detonation wave surface and the second high-pressure region are connected together. Subgraph 1 is the local amplification of the pressure peak of the detonation wave. The three pressure distribution curves basically coincide, which indicates that the detonation wave can be well captured when the grid size is 0.05-0.15 mm. When the grid size is 0.05 mm, 0.1 mm and 0.15 mm, the detonation wave velocity is 1970 m/s, 1970 m/s and 1964.8 m/s respectively. This indicates that the detonation wave velocity reaches grid convergence, and these three velocities are very close to the theoretical C-J velocity of the detonation wave, which is 1971.5 m/s.

The working conditions corresponding to the case in Fig. 2(b) are as follows: the hydrogen equivalence ratio is 0.7, the carbon particle equivalence ratio is 1.11, the carbon particle radius is 2 μm, the oxygen mass fraction in the oxygen-rich air is 0.283, the initial pressure of the reactant is 1atm, and the initial temperature is 300 K. As shown in the figure, the detonation wave in this case has two pressure peaks, that is, the gas phase detonation wave surface and the second high-pressure region are separated. Subgraph 2 is the local amplification of the pressure peak of the gas phase detonation wave, and subgraph 3 is the local amplification of the pressure peak of the second high-pressure region. The three pressure distribution curves basically coincide, which indicates that the detonation wave can be well captured when the grid size is 0.05-0.15 mm. When the grid size is 0.05 mm, 0.1 mm and 0.15 mm, the detonation wave velocity is 1850 m/s, 1850 m/s and 1844.8 m/s respectively. This indicates that the detonation wave velocity reaches grid convergence.

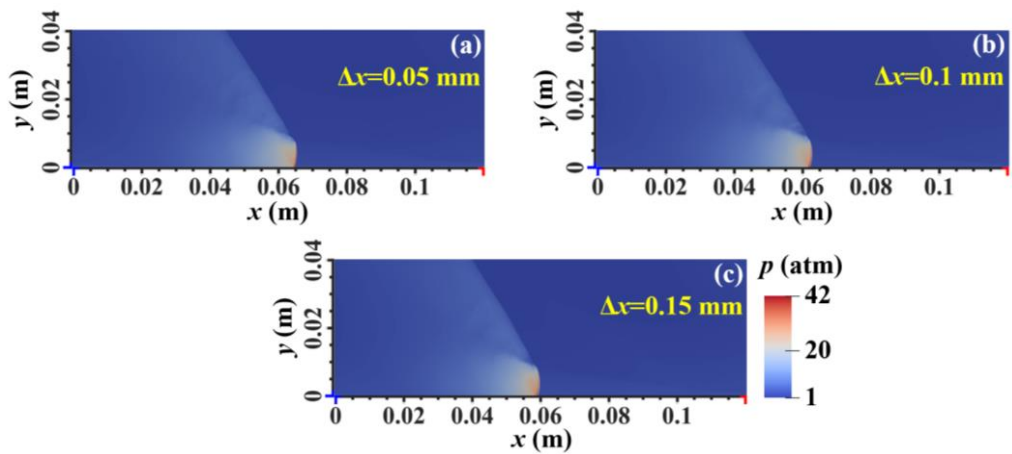

**Fig. 3 Pressure nephogram of two-dimensional RDE flow field.**

Next, the two-dimensional RDE with different grid sizes is numerically simulated to verify the grid independence. The working conditions corresponding to the case in Fig. 3 are as follows: the hydrogen equivalent ratio is 0.5, the carbon particle equivalent ratio is 2.29, the carbon particle radius is 1 μm, the velocity of carbon particles injected into RDE is 50 m/s, the oxygen mass fraction in oxygen-rich air is 0.457, the injection total pressure of gas phase is 4 atm, and the injection total temperature is 360 K. The corresponding grid sizes of Fig. 3(a)-(c) are 0.05 mm, 0.1 mm and 0.15 mm, respectively. Under different grid sizes, there is one rotating detonation wave in the RDE flow field, and the flow field and wave structure are basically the same. This shows that for the cases with three grid sizes, the captured wave structures are exactly the same, only the resolutions are different. Fig. 2 and Fig. 3 verify the grid independence. In order to reduce the calculation cost under the premise of ensuring the calculation accuracy, the grid size selected in this study is 0.1 mm. According to the error estimation method proposed by Smirnov et al. [70,71], the total error of the case in this study is less than 3%.

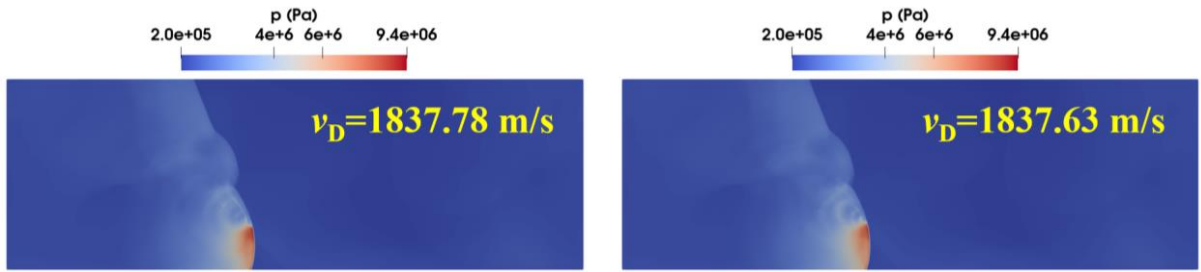

Fig. 4 Flow field without considering pressure gradient force (left) and considering pressure gradient force (right).

The following investigates the flow fields of RDE without considering or considering the pressure gradient force. Fig. 4 shows the flow fields without considering the pressure gradient force (left) and with considering the pressure gradient force (right). In the left figure, the force acting on solid particles is represented by Eq. 14, while in the right figure, the force acting on solid particles is represented by Eq. 20. In Eq. 20, the first term on the right-hand side represents the drag force, and the second term represents the pressure gradient force. By comparing the two results, it can be observed that the flow field and wave structure remain the same, and the detonation wave velocities are 1837.78 m/s and 1837.63 m/s for the cases without considering and with considering the pressure gradient force, respectively. The error in the detonation wave velocity when not considering the pressure gradient force is 0.00816%, which is very small.

$$F_p = m_p \frac{u - u_p}{\tau_r} + m_p \frac{\rho}{\rho_p} u_p \nabla \cdot u \tag{20}$$

## 3. Results and discussion

### 3.1 Flow field characteristics and wave structure of gas-solid two-phase RDE

#### 3.1.1 Gas-solid double-front detonation structure

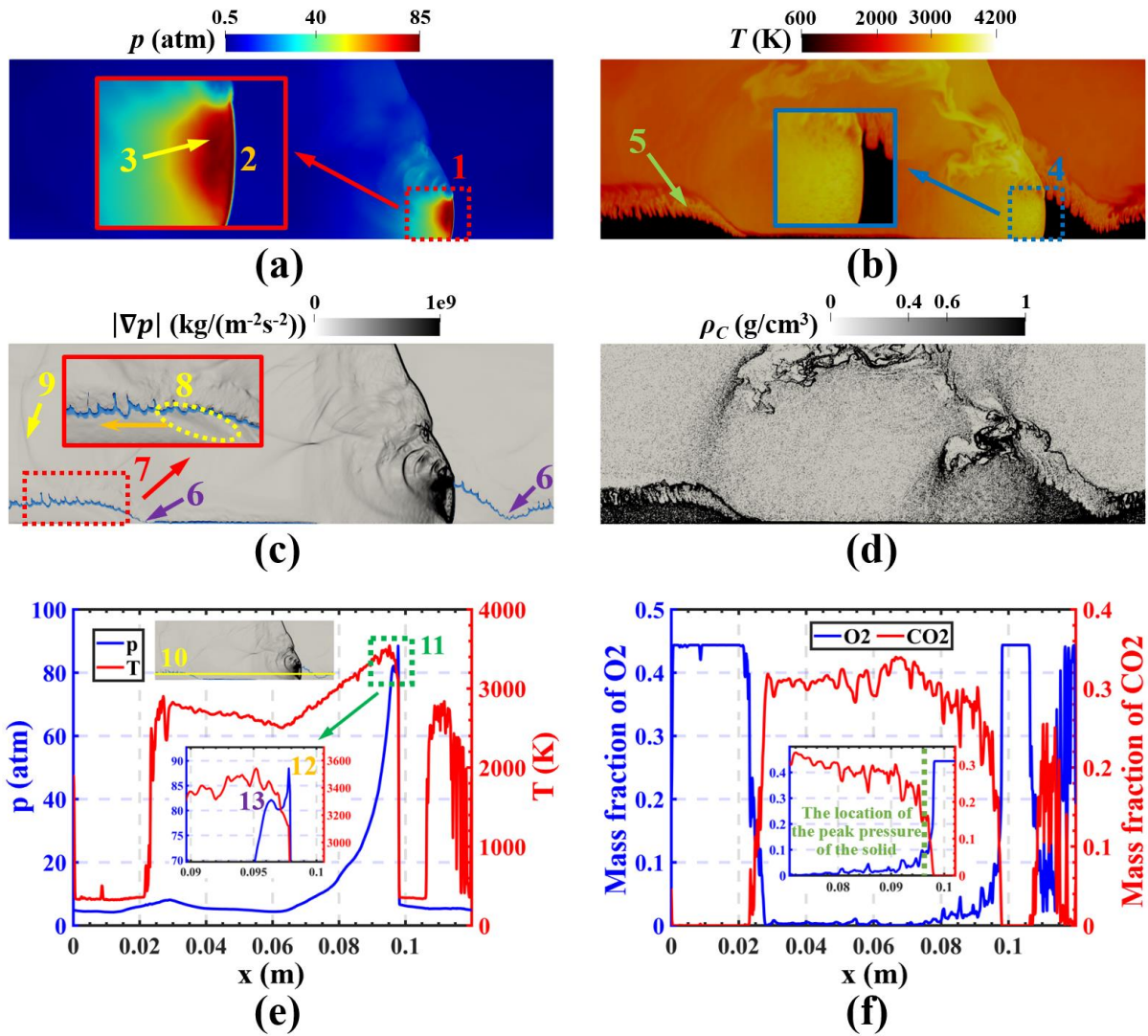

**Fig. 5 Physical quantity distribution and component mass fraction nephogram of the flow field of Case 3. (a) Pressure nephogram. (b) Temperature nephogram. (c) Pressure gradient nephogram. (d) Solid particle density nephogram. (e) The curves of pressure and temperature changing with circumferential position. (f) The curves of oxygen mass fraction and carbon dioxide mass fraction change with circumferential position.**

Fig. 5 shows the distribution of physical quantities and component mass fractions in the flow field of Case 3. The figure illustrates the flow field characteristics of a detonation wave when the solid particle diameter is 1 μm. In Fig. 5(a), the pressure nephogram reveals the presence of a detonation wave in the flow field, with the tail of the detonation wave connected to an oblique shock wave. The local amplification of the detonation wave head is highlighted by the red box 1, which shows two detonation wave surfaces. The gaseous fuel exhibits a high chemical reaction rate, placing the gaseous detonation wave front ahead and marked with the orange number 2. Conversely, the solid fuel has a low chemical reaction rate, causing the second high-pressure region to lag behind, marked by yellow arrow 3. This structure represents a typical gas-solid two-phase mixed double-front detonation structure.

In Fig. 5(b), the temperature nephogram indicates that the temperature behind the gas-phase detonation wave surface continues to rise. This increase is attributed to the ongoing combustion and heat release of the solid-phase fuel after the gas-phase detonation wave. The black area represents fresh gas, and the interface between fresh gas and combustion products exhibits an irregular, jagged shape. This instability at the interface is enhanced by the combustion of hydrogen, as indicated by green arrow 5, which also accelerates the solid phase combustion.

Fig. 5(c) presents the pressure gradient nephogram of the flow field. The light blue boundary line represents the fresh gas layer, which displays two characteristics. Firstly, it exhibits an overall jagged instability, and secondly, it shows local depressions, as indicated by purple arrow 6. The depression of fresh gas is caused by the counter-rotating shock wave obstructing the inflow of fresh gas. The local amplification of the counter-rotating shock wave is depicted by the red box 7. The counter-rotating shock wave moves in the opposite direction to the detonation wave, with the part within the fresh gas represented by the yellow ellipse 8, and the part outside the fresh gas indicated by the yellow arrow 9. The quantitative analysis of the formation of counter-rotating shock waves is given in Appendix 2. Fig. 5(d) shows the solid particle density nephogram of the flow field, illustrating the distribution of solid fuel. Clearly, the solid particles are primarily concentrated in the fresh gas, while their density significantly decreases after the detonation wave.

Fig. 5(e)-(f) presents the distribution of physical quantities near the inlet in the flow field of Case 3, displaying variations with circumferential position. The yellow line 10 indicates the position of the collected data. By analyzing the data curve, one can obtain accurate distribution characteristics of each physical quantity near the detonation wave head. In Fig. 5(e), the curve illustrates the changes in pressure and temperature with circumferential position. The blue curve represents pressure, the red curve represents temperature, and the green box 11 provides a local amplification of the gas-solid two-phase double-front detonation structure. Within the pressure curve, the gas-phase detonation wave surface is identified by the orange number 12, while the second high-pressure region is indicated by the purple number 13. The pressure peak of the gas-phase detonation wave surface is higher than that of the second high-pressure region. Regarding the temperature curve, there is a sharp rise at the gas-phase detonation wave surface, followed by a gradual increase behind the gas-phase wave surface. In Fig. 5(f), the curve represents the changes in oxygen mass fraction and carbon dioxide mass fraction with circumferential position. The local magnification in the figure highlights the gradual decrease of oxygen and increase of carbon dioxide after the hydrogen detonation wave surface.

### 3.1.2 Gas-solid two-phase single-front detonation structure

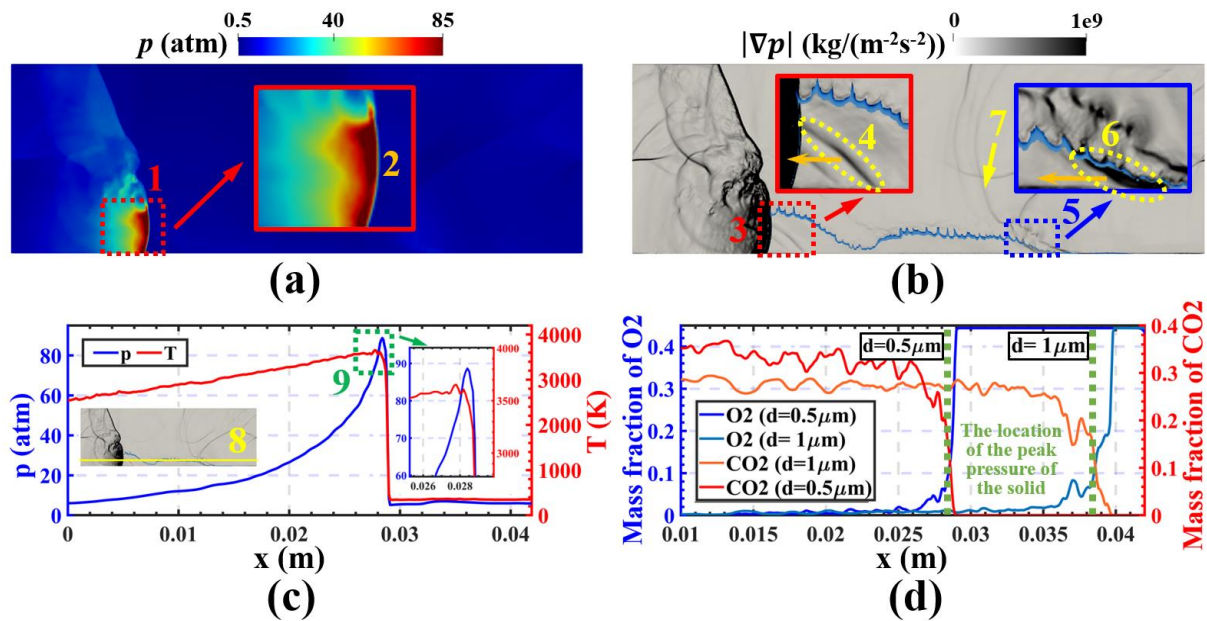

**Fig. 6 Physical quantity distribution nephogram and physical quantity-position curve of the flow field of Case 2. (a) Pressure nephogram. (b) Pressure gradient nephogram. (c) The curves of pressure and temperature changing with circumferential position. (d) The curves of oxygen mass fraction and carbon dioxide mass fraction change with circumferential position.**

Fig. 6 shows the physical quantities distribution nephogram, and physical quantity-position curve for Case 2, illustrating the flow field characteristics of a detonation wave with a solid particle diameter of 0.5 μm. In Fig. 6(a), the pressure nephogram shows the flow field's detonation wave head, with the red box 1 indicating a local amplification. Unlike the case with a solid particle diameter of 1 μm, Case 2 exhibits only one detonation wave surface at its detonation wave head, identified by the orange number 2. This difference arises from reducing the solid particle diameter, which increases the contact area between the solid fuel and oxidant, resulting in a higher chemical reaction rate. Consequently, there is a more intense chemical reaction of the solid fuel and a connection between the second high-pressure region and the gas-phase detonation wave surface. This configuration represents a gas-solid two-phase mixed single-front detonation structure.

Fig. 6(b) shows the pressure gradient nephogram of the flow field. The light blue boundary line represents the fresh gas layer, which exhibits local depressions caused by the blocking of fresh gas by the counter-rotating shock wave. The red box 3 and blue box 5 provide local amplifications of the counter-rotating shock wave. The colorbar range in the amplification diagram is set as [0, 6e7] to clearly display the counter-rotating shock wave. The flow field contains two counter-rotating shock waves, represented by the yellow ellipses 4 and 6 within the fresh gas. The orange arrow indicates the direction of the counter-rotating shock wave, which is opposite to the direction of the detonation wave. The yellow arrow 7 represents the part of the counter-rotating shock wave outside the fresh gas.

Fig. 6(c) shows the curve of pressure and temperature with respect to circumferential position. The yellow line 8 represents the position of the collected data. The blue curve represents pressure, the red curve represents temperature, and the green box 9 provides a local amplification of the gas-solid two-phase single-front detonation structure. The pressure curve exhibits a single peak, and the temperature and pressure reach their peaks almost simultaneously. However, both pressure and temperature curves show two rising sections with different slopes, indicating that although the gas-phase detonation wave surface and the solid-phase high-pressure region are in contact, they do not completely coincide.

Fig. 6(d) presents the curve of oxygen mass fraction and carbon dioxide mass fraction with respect to circumferential position. The curves for solid particle diameters of 0.5 μm and 1 μm are plotted for comparison. After analysis, it is evident that the oxygen reduction rate and carbon dioxide increase rate in the flow field with a solid particle diameter of 0.5 μm are significantly higher than those in the flow field with a solid particle diameter of 1 μm. This indicates that decreasing the solid particle diameter increases the chemical reaction rate of the solid fuel.

### 3.1.3 Gas-phase single-front detonation structure

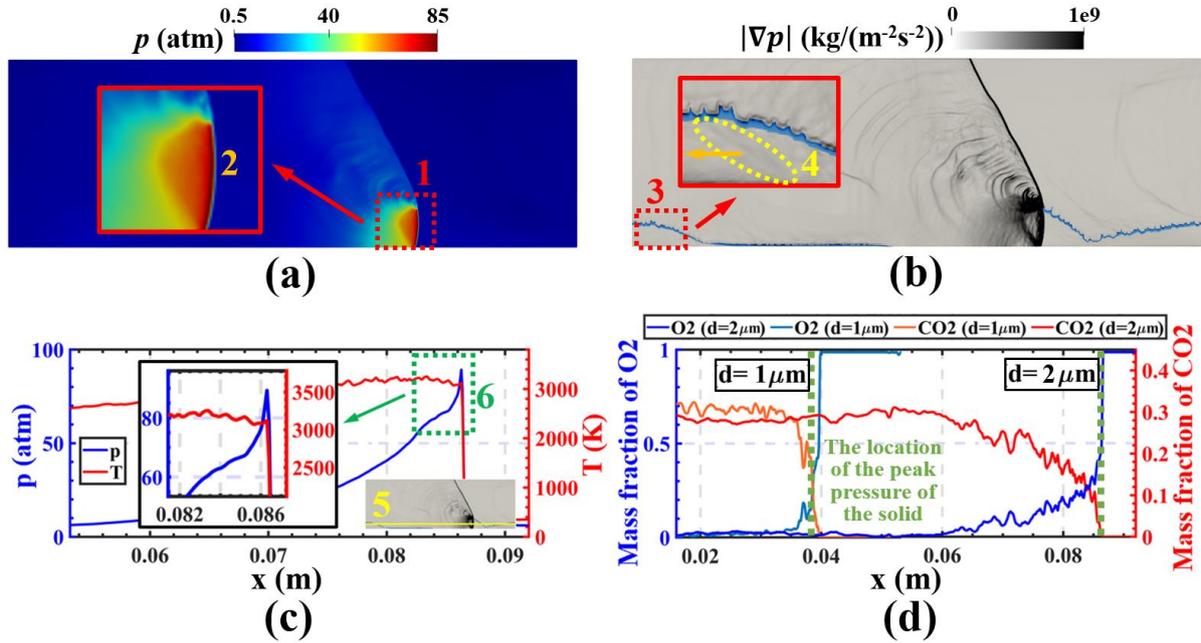

**Fig. 7 Physical quantity distribution nephogram and physical quantity-position curve of the flow field of Case 4. (a) Pressure nephogram. (b) Pressure gradient nephogram. (c) The curves of pressure and temperature changing with circumferential position. (d) The curves of oxygen mass fraction and carbon dioxide mass fraction change with circumferential position.**

Fig. 7 illustrates the flow field physical quantities and component distribution nephogram, along with the physical quantity-position curve of Case 4, showing the flow field characteristics of the rotating detonation wave with a solid particle diameter of 2 μm. In Fig. 7(a), the pressure nephogram of the flow field is presented. The red box 1 denotes the local amplification of the detonation wave head, and the single detonation wave surface is identified by the orange number 2. Unlike Case 2, Case 4 exhibits a gas-phase detonation wave surface as the only detonation wave, while the solid fuel combusts slowly behind the detonation wave without forming a second high-pressure region. This phenomenon occurs due to the increased diameter of the solid particles, which decreases the surface area of the solid fuel, leading to a reduced chemical reaction rate and slower reaction of the solid fuel. The resulting structure is a gas-phase single-front detonation structure.

Fig. 7(b) shows the pressure gradient nephogram of the flow field, where the light blue boundary line represents the fresh gas layer. The fresh gas layer exhibits a local depression caused by the counter-rotating shock wave. The red box 3 represents the local amplification of the counter-rotating shock wave, the yellow ellipse 4 indicates the portion of the counter-rotating shock wave within the fresh gas, and the orange arrow indicates the direction of the counter-rotating shock wave.

Fig. 7(c) shows the curves of pressure and temperature changing with circumferential position. The yellow line 5 denotes the position of the collected data, while the green box 6 represents the local amplification of the gas-phase single-front detonation structure. In the pressure curve, only one gas-phase peak is observed, and the temperature shows slight fluctuations with minimal increase behind the gas-phase peak.

Finally, Fig. 7(d) exhibits the curve of oxygen mass fraction and carbon dioxide mass fraction changing with circumferential position. The curves corresponding to solid particle diameters of 2 μm and 1 μm are presented for comparison. Upon analysis, it is evident that the oxygen reduction rate and carbon dioxide increase rate in the flow field with a solid particle diameter of 2 μm are significantly smaller than those in the flow field with a solid particle diameter of 1 μm.

**3.1.4 The comparison and summary of the three structures**

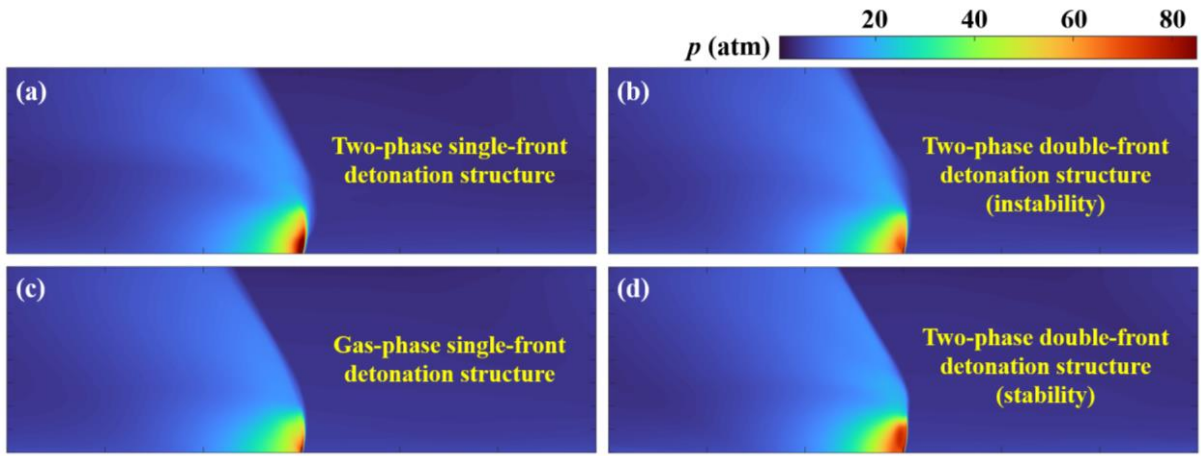

**Fig. 8 The time-averaged fields extracted relative to the detonation wave reference frame, This will allow the wave to be frozen in the middle of the domain. (a) Case 2, two-phase single-front detonation structure. (b) Case 3, two-phase double-front detonation structure (unstable detonation flow field). (c) Case 4, gas-phase single-front detonation structure. (d) Case 2, two-phase double-front detonation structure (strongly stable detonation flow field).**

Fig. 8 shows the time-averaged fields extracted relative to the detonation wave reference frame. This will allow the wave to be frozen in the middle of the domain. The time-averaged fields obtained using this method provide detailed information about the detonation structures. Fig. 8 (a)-(c) display the time-averaged fields for three different detonation structures, and below is a comparative analysis of these three structures. In Fig. 8 (a), for the gas-solid two-phase single-front detonation structure, the gas-phase detonation wave is connected to the high-pressure region of solid particles behind it. Compared to the other structures, the pressure in the solid-phase high-pressure region is significantly higher in this structure.

In Fig. 8 (b), for the gas-solid two-phase double-front detonation structure, the gas-phase detonation wave propagates forward, and solid particles burn behind the detonation wave, forming a secondary high-pressure region. Due to the instability of the detonation wave, there are fluctuations in its height, which leads to fluctuations in the height of the solid-phase high-pressure region. As a result, the time-averaged pressure in the unstable solid-phase high-pressure region is lower than the maximum instantaneous pressure observed in the pressure field. Comparing Fig. 8 (d) with Fig. 8 (b), Fig. 8 (d) displays the time-averaged field of the two-phase double-front detonation structure with strong detonation flow field stability, and it can be observed that the solid-phase high-pressure region is quite prominent. However, the pressure in the solid-phase high-pressure region in both time-averaged fields is lower than that in the solid-phase high-pressure region of the gas-solid two-phase single-front detonation structure.

In Fig. 8 (c), for the gas-phase single-front detonation structure, solid particles burn slowly behind the gas-phase detonation wave, and there is no formation of a solid-phase high-pressure region behind the detonation wave.

The reason for these differences is that, at the same solid equivalence ratio, reducing the diameter of solid particles increases the total surface area of the solid phase, thereby accelerating the reaction rate of solid particles. In the case of gas-solid two-phase single-front structure, the solid particle diameter is the smallest (0.5 μm), resulting in the fastest reaction rate of solid particles and no significant distance between the gas-phase detonation wave and the solid-phase high-pressure region. In the gas-solid two-phase double-front structure, the solid particle diameter is moderate (1 μm), resulting in a relatively fast reaction rate of solid particles and a certain distance between the gas-phase detonation wave and the solid-phase high-pressure region. In the case of gas-phase single-front structure, the solid particle diameter is the largest (2 μm), resulting in the slowest reaction rate of solid particles, and no solid-phase high-pressure region is formed in the flow field.

It is worth noting that the time-averaged fields computed and extracted relative to the detonation wave reference frame can provide both detailed information about the detonation structures and quantitative information for flow field analysis. In the subsequent quantitative analysis of the transition process between different detonation modes, the data of characteristic physical quantities are obtained through time-averaged fields. Therefore, the time-averaged fields obtained relative to the detonation wave reference frame hold significant importance.

**3.2 The impact of particle diameter and hydrogen equivalence ratio on the flow field**

**3.2.1 The influence of particle diameter and hydrogen equivalence ratio on detonation modes and self-sustained propagation**

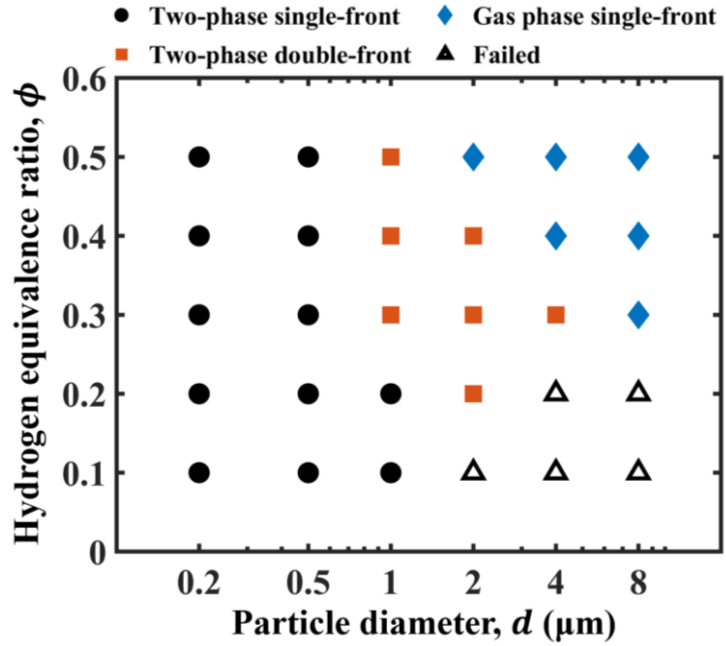

**Fig. 9 The case operating modes (i.e., two-phase single-front detonation, two-phase double-front detonation, gas-phase single-front detonation, or failure). The vertical axis represents the hydrogen equivalence ratio, and the horizontal axis represents the particle diameter.**

Fig. 9 shows the ability of detonation waves to self-sustain and their detonation modes under different particle diameters and hydrogen equivalence ratios. Solid black circles represent rotating detonation waves with a gas-solid two-phase single-front structure, solid red squares represent rotating detonation waves with a gas-solid two-phase double-front structure, solid blue diamonds represent rotating detonation waves with a gas-phase single-front structure, and hollow black triangles indicate detonation waves that cannot self-sustain propagation.

When the hydrogen equivalence ratio ranges from 0.3 to 0.5, the detonation wave can self-sustain propagation, and as the particle diameter increases, the detonation wave transitions from a gas-solid two-phase single-front structure to a gas-solid two-phase double-front structure, and then to a gas-phase single-front structure. The difference is that as the hydrogen equivalence ratio decreases, the range of particle diameters that can achieve a gas-solid two-phase double-front structure increases. From the operation mode chart, it can be observed that the cases exhibiting gas-phase single-front detonation are located in the upper right corner of the chart, indicating that reducing the particle diameter and hydrogen equivalence ratio (not less than 0.3) contributes to the realization of gas-solid two-phase rotating detonation.

When the hydrogen equivalence ratio ranges from 0.1 to 0.2, as the hydrogen equivalence ratio decreases, the range of particle diameters that can achieve self-sustained propagation of the detonation wave becomes smaller. From the operation mode chart, it can be observed that the cases of detonation failure are located in the lower right corner of the chart, indicating that reducing the particle diameter and increasing the hydrogen equivalence ratio help to avoid detonation wave extinction. Within this range of hydrogen equivalence ratio, the range of particle diameters that can form a gas-solid two-phase single-front detonation increases, suggesting that reducing the hydrogen equivalence ratio contributes to the realization of gas-solid two-phase single-front rotating detonation.

In summary, reducing both the hydrogen equivalence ratio and the particle diameter contribute to the transition from a gas-phase single-front detonation to a gas-solid two-phase double-front detonation, and then to a gas-solid two-phase single-front detonation. This is because decreasing the hydrogen equivalence ratio implies more solid particles undergoing reactions, and reducing the particle diameter increases the particle's specific surface area and reaction rate, both of which promote the formation and enhancement of high-pressure regions in the solid phase. When the hydrogen equivalence ratio becomes too small and the particle diameter of the solid fuel becomes too large, the detonation wave cannot sustain self-propagation. This is due to the reduction in hydrogen content and the decrease in specific surface area of the particles, which lowers the reaction rate and leads to detonation failure.

**3.2.2 Quantitative analysis of transitions of the rotating detonation mode**

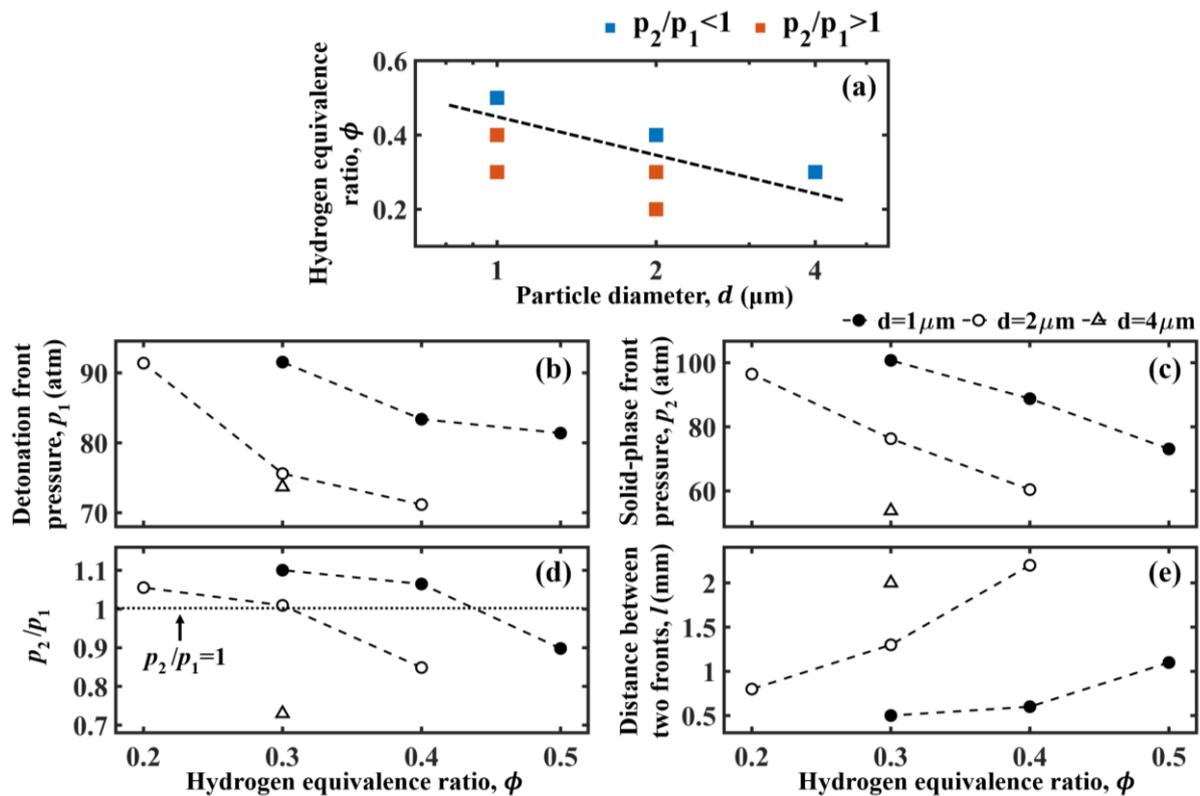

Fig. 10 Variation of characteristic physical quantities of the gas-solid two-phase double-front structure with respect to particle diameter and hydrogen equivalence ratio. (a) The case operating modes (indicated by $p_2/p_1<1$ and $p_2/p_1>1$). (b) Variation of the pressure at the detonation front, $p_1$. (c) Variation of the pressure at the solid phase front, $p_2$. (d) Variation of the ratio of the pressure at the solid phase front to the pressure at the detonation front, $p_2/p_1$. (e) Variation of the distance between the two fronts, *l*.

Next, we study the transitional dynamics of the rotating detonation mode. The two-phase double-front structure is an intermediate mode that transits from a two-phase single-front structure to a gas-phase single-front structure. In Section 3.1, we analyzed the characteristics of the two-phase double-front detonation structure, which includes a detonation front and a solid-phase high-pressure region, exhibiting two pressure peaks in the pressure-position curve. To clarify the variation of the wave system structure of the gas-solid two-phase double-front with respect to particle diameter and hydrogen equivalence ratio, it is necessary to conduct a quantitative analysis of the characteristic physical quantities of the double-front wave system. These quantities include the pressure at the detonation front, $p_1$, the pressure at the solid-phase front, $p_2$, the ratio of the pressure at the solid-phase front to the pressure at the detonation front, $p_2/p_1$ (which reflects the intensity of the solid-phase high-pressure region), and the distance between the two fronts, *l*.

Fig. 10(a) presents the operating conditions for the cases that achieved the two-phase double-front detonation in Fig. 9. The cases where the pressure at the solid-phase front is greater than the pressure at the detonation front ($p_2/p_1>1$) are represented by red squares, and the cases where the pressure at the solid-phase front is less than the pressure at the detonation front ($p_2/p_1<1$) are represented by blue squares. The cases with smaller pressure at the solid-phase front are concentrated in the upper-right region of the operating mode diagram, indicating that reducing the particle diameter and hydrogen equivalence ratio contribute to increasing the intensity of the solid-phase high-pressure region.

Figs. 10(b)-(e) show the variation of characteristic physical quantities for the 7 cases presented in Fig. 10(a) with respect to particle diameter and hydrogen equivalence ratio. Through analysis, it can be observed that the pressure at the detonation front, $p_1$, the pressure at the solid-phase front, $p_2$, and the ratio of $p_2/p_1$ all decrease with increasing hydrogen equivalence ratio and increasing particle diameter. On the other hand, the distance between the two fronts, *l*, increases with increasing hydrogen equivalence ratio and increasing particle diameter. Based on the analysis mentioned earlier, reducing the particle diameter can increase the reaction rate of the solid-phase, and decreasing the hydrogen equivalence ratio can increase the mass of the solid-phase participating in the reaction. Both of these factors contribute to increasing the intensity of the solid-phase high-pressure region.

**3.2.3 The influence of particle diameter and hydrogen equivalence ratio on the rotating detonation wave speed and stability**

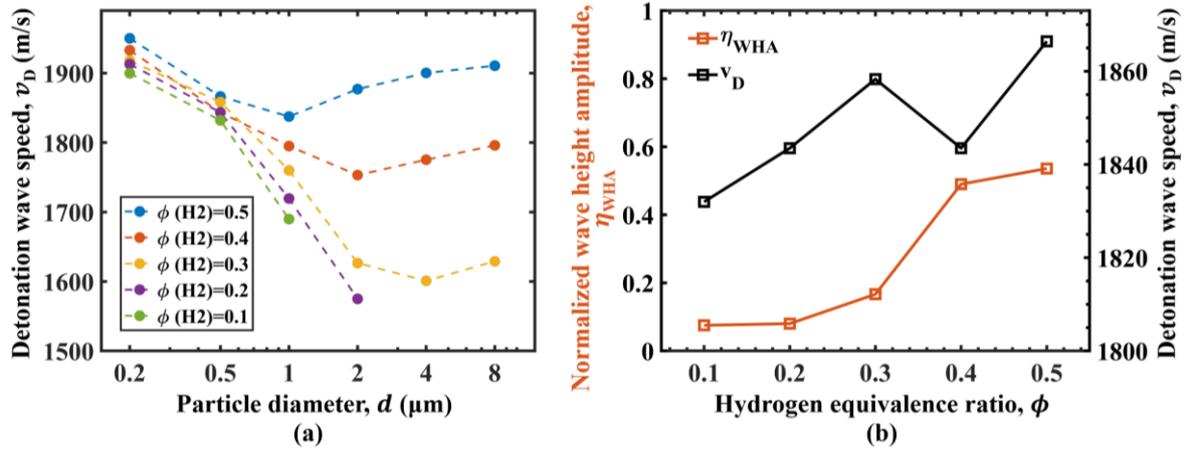

Fig. 11 (a) The variation of detonation wave speed with particle diameter and hydrogen equivalence ratio. (b) At a particle diameter of 0.5 μm, the variation of detonation wave speed and stability with hydrogen equivalence ratio, to explain the mechanism behind the non-monotonic change of detonation wave speed under this condition.

The propagation characteristics and stability of rotating detonation waves directly determine the smooth operation of rotating detonation engines. Therefore, the speed and stability of rotating detonation waves are important indicators for studying the flow field characteristics of RDEs. Fig. 11(a) shows the variation of detonation wave speed with particle diameter and hydrogen equivalence ratio. Firstly, the effect of particle diameter on the detonation wave speed is analyzed. When the hydrogen equivalence ratio is in the range of 0.3-0.5, the detonation wave speed first decreases and then increases. Comparing Fig. 9 and Fig. 11(a), the minimum value of detonation wave speed corresponds to the detonation mode of two-phase double-front structure. When the detonation wave exhibits two-phase single/double-front structure, the detonation wave speed increases with the decrease in particle diameter. This is because reducing the particle diameter increases the solid phase surface area, thus enhancing the chemical reaction rate and ultimately increasing the strength and speed of the two-phase detonation wave.

When the solid particles are relatively large and only gas phase detonation waves exist, the detonation wave speed increases with the increase in particle diameter. This is because larger solid particles have smaller surface areas, resulting in less heat transfer from gas phase detonation waves to solid particles. As a result, the detonation wave speed and strength become higher.

Now the effect of hydrogen equivalence ratio on the detonation wave speed is analyzed. When the particle diameter is 0.2 μm and 1-8 μm, the detonation wave speed decreases with the decrease in hydrogen equivalence ratio. However, when the particle diameter is 0.5 μm, there is a phenomenon where the detonation wave speed does not vary monotonically with the equivalence ratio, as shown by the black line in Fig. 11(b). Now, the mechanism behind this phenomenon is analyzed.

Based on the analysis in Section 3.1, under certain operating conditions, fresh gas ahead of the detonation wave can undergo a concave shape, leading to fluctuations in the height of the detonation wave front. We define the normalized wave height amplitude to quantify the instability of the detonation wave:

$$\eta_{\text{WHA}} = \frac{\dfrac{h_{\max} - h_{\min}}{2}}{\dfrac{h_{\max} + h_{\min}}{2}} = \frac{h_{\max} - h_{\min}}{h_{\max} + h_{\min}} \quad (21)$$

The parameter $\eta_{\text{WHA}}$ represents the normalized wave height amplitude, where $h_{\max}$ denotes the maximum wave height and $h_{\min}$ denotes the minimum wave height. The physical meaning of $\eta_{\text{WHA}}$ is the ratio of the amplitude of detonation wave front height fluctuations to the average height of the detonation wave front. A higher $\eta_{\text{WHA}}$ indicates a stronger instability in the propagation process of the detonation wave.

In Fig. 11(b), the red line represents the variation of $\eta_{\text{WHA}}$ with the hydrogen equivalence ratio for a particle diameter of 0.5 μm, indicating that the detonation instability decreases with a decrease in the hydrogen equivalence ratio. This phenomenon occurs because gaseous fuels burn more vigorously than solid fuels. When the proportion of gaseous fuel decreases, the shock waves that cause instability in the flow field will also be reduced and weakened, resulting in less impact on the fresh gas ahead of the detonation wave. Thus, the detonation instability is reduced significantly as the hydrogen equivalence ratio decreases from 0.4 to 0.3, as indicated by the decrease in $\eta_{\text{WHA}}$ from 0.490 to 0.167. When the detonation wave becomes more stable, its strength and speed increase, leading to occasional cases where the detonation wave speed does not vary monotonically with the equivalence ratio. In summary, in most cases, the detonation wave speed decreases with a decrease in the hydrogen equivalence ratio. Yet, due to variations in detonation stability, the detonation wave speed can exhibit non-monotonic behavior.

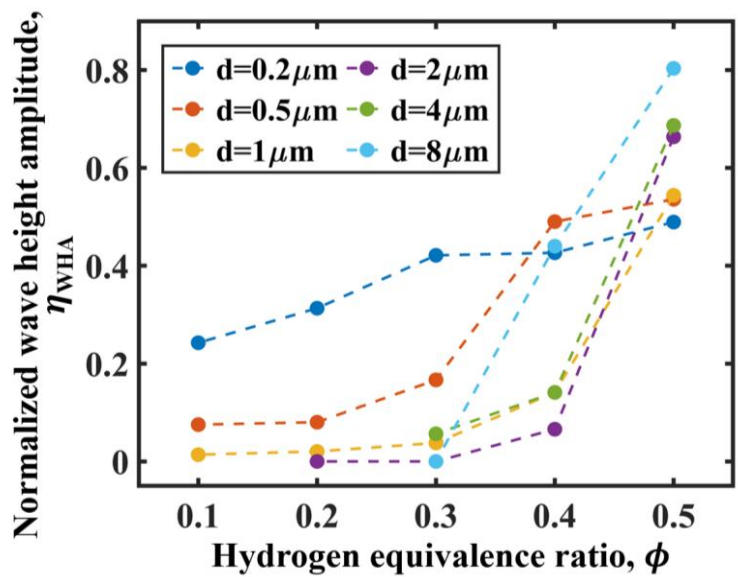

**Fig. 12 The stability of the detonation wave varies with the particle diameter and hydrogen equivalence ratio. The stability is characterized using the normalized wave height amplitude, $\eta_{WHA}$.**

Now the stability of the detonation wave with respect to the particle diameter and hydrogen equivalence ratio is analyzed. As mentioned earlier, it was observed that when the particle diameter is 0.5 μm, the detonation instability decreases with a decrease in the hydrogen equivalence ratio, and the reasons were discussed. Fig. 12 shows the variation of detonation wave stability with the hydrogen equivalence ratio for particle diameters ranging from 0.2 μm to 8 μm. It can be observed that in all cases, the detonation instability decreases with a decrease in the hydrogen equivalence ratio.

Next, the effect of particle diameter on the instability of the detonation wave is analyzed. When the hydrogen equivalence ratio is in the range of 0.1-0.2, where the self-sustained propagation of the detonation wave mainly relies on the solid phase, the instability of the detonation wave increases with a decrease in the particle diameter. This is because a smaller particle diameter leads to an increase in the reaction rate, resulting in vigorous combustion, which generates more and stronger shock waves in the flow field. These shock waves have a greater impact on the fresh gas ahead of the detonation wave, ultimately enhancing the instability of the detonation wave.

### 3.3 Performance analysis

### 3.3.1 Pressure gain characteristic analysis

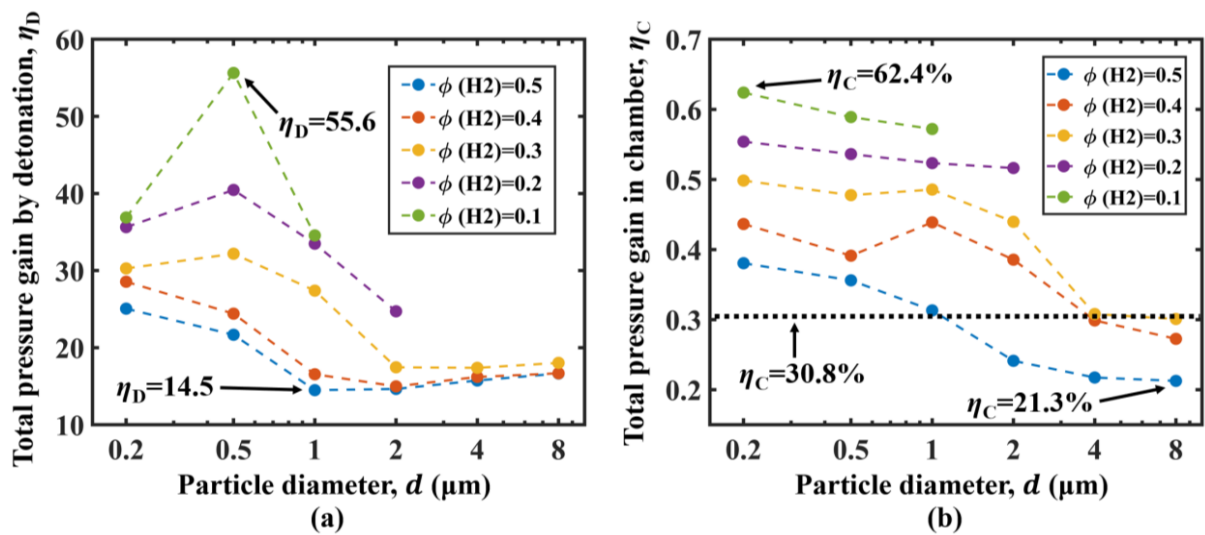

**Fig. 13 The total pressure gain of the detonation wave $\eta_D$ and the total pressure gain of the rotating detonation combustor $\eta_C$ vary with the particle diameter and hydrogen equivalence ratio.**

Rotating detonation engines exhibit the characteristic of self-pressurization, which converts more thermal energy into kinetic energy. Due to the challenges in directly measuring the total pressure at the rotating detonation combustor and its exit through experiments, it is necessary to use numerical simulations to reveal the pressure gain characteristics of the gas-solid two-phase rotating detonation engine. The total pressure gain of the rotating detonation wave $\eta_D$ and the total pressure gain of the combustor $\eta_C$ are analyzed below to evaluate the pressure gain performance of the gas-solid two-phase rotating detonation. Fig. 13(a) shows the variation of the total pressure gain $\eta_D$ of the detonation wave with particle diameter and hydrogen equivalence ratio. The expression for $\eta_D$ is as follows:

$$\eta_D = \frac{\overline{p_D} - p_3}{\overline{p_3}} \tag{22}$$

Where $\overline{p_D}$ is the average total pressure of the detonation wave, and $p_3$ is the total pressure ahead of the detonation wave. By comparing and analyzing the results of all cases, it is found that as the hydrogen equivalence ratio decreases, the total pressure gain of the detonation wave $\eta_D$ gradually increases and ranges between 14.5 and 55.6. This indicates that the rotating detonation wave serves as a source of pressure gain for the rotating detonation engine. Fig. 13(b) shows the variation of the total pressure gain $\eta_C$ of the combustor with particle diameter and hydrogen equivalence ratio. The expression for $\eta_C$ is as follows:

$$\eta_C = \frac{\overline{p_C} - p_0}{\overline{p_0}} \tag{23}$$

Among them, $\overline{p_C}$ is the average total pressure at the exit of the combustor, and $p_0$ is the injection total pressure. By comparing and analyzing the results of all cases, it is found that as the hydrogen equivalence ratio decreases, the total pressure gain of the combustor $\eta_C$ gradually increases. When the detonation wave exhibits a two-phase single/double-front structure, $\eta_C$ is greater than 30.8%; when the detonation wave exhibits a gas-phase single-front structure, $\eta_C$ is less than 30.8%. That is, the $\eta_C$ corresponding to the two-phase detonation wave is greater than that of the gas-phase detonation wave. In most cases, as the particle diameter decreases, the total pressure gain of the combustor $\eta_C$ gradually increases. This is because the decrease of particle diameter increases the reaction rate, increases the solid mass that participates in the reaction and contributes to the formation of detonation front and solid front, ultimately leading to an increase in the overall total pressure gain of the combustor.

However, when the hydrogen equivalence ratio is 0.4 and 0.3, and the particle diameter decreases from 1 μm to 0.5 μm, the total pressure gain $\eta_C$ of the combustor decreases. This is due to changes in the instability of the detonation. As shown in Table 3, when the hydrogen equivalence ratio is 0.3 and the particle diameter decreases from 1 μm to 0.5 μm, $\eta_{WHA}$ increases from 0.038 to 0.166; when the hydrogen equivalence ratio is 0.4 and the particle diameter decreases from 1 μm to 0.5 μm, $\eta_{WHA}$ increases from 0.141 to 0.490. When $\eta_{WHA}$ increases, indicating an enhancement in the instability of the rotating detonation flow field, more deflagration is produced in the flow field, which does not contribute to the self-pressure gain effect and does not contribute to the total pressure gain of the combustor.

In summary, the total pressure gain of the combustor increases as the hydrogen equivalence ratio decreases, and the increase in the mass of solid fuel participating in detonation combustion significantly raises the total pressure gain of the combustor. Generally, the total pressure gain of the combustor decreases with an increase in the particle diameter. However, this trend can also be affected by the instability of the detonation flow field. The enhancement of detonation instability can lead to a loss in the total pressure gain of the combustor.

Table 3 The normalized wave height amplitude ($\eta_{WHA}$) for different hydrogen equivalence ratios and particle diameters in the cases.

| $\phi(H_2)$ \ $d$ (μm) | 0.5 | 1 |
|---|---|---|
| 0.3 | 0.166 | 0.038 |
| 0.4 | 0.490 | 0.141 |

### 3.3.2 Propulsion performance

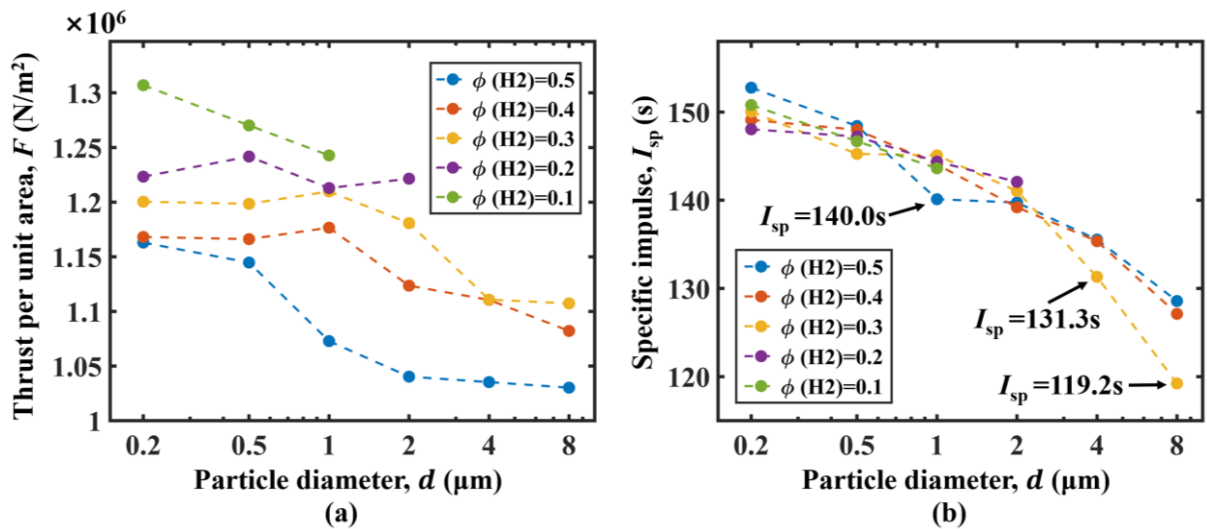

Fig. 14 Thrust per unit area ($F$) and specific impulse ($I_{sp}$) as a function of particle diameter and hydrogen equivalence ratio. (a) Thrust per unit area ($F$). (b) Specific impulse ($I_{sp}$).

Thrust and specific impulse are important parameters for evaluating the propulsion performance of the gas-solid two-phase rotating detonation engine, where specific impulse measures the propellant consumption when generating thrust. In this section, the thrust and specific impulse of cases with different particle diameters and hydrogen equivalence ratios were calculated and compared. The thrust per unit area of the combustion chamber and the unit area mass flow rate of the propellant are calculated as follows:

$$F = \frac{\oint_{outlet} [\rho v^2 + p - p_\infty] r d\theta}{2\pi r} \quad (24)$$

$$\dot{m} = \frac{\oint_{inlet} [\rho v] r d\theta}{2\pi r} + \dot{m}_s \quad (25)$$

Where $p_\infty$ is the ambient pressure, equal to 1 atm, $\rho$ is the gas density, $v$ is the axial velocity, $p$ is the pressure, $r$ is the combustion chamber radius, and $\dot{m}_s$ represents the unit area mass flow rate of the solid fuel. Based on the thrust per unit area ($F$) and the unit area mass flow rate of the propellant ($\dot{m}$), the specific impulse is calculated as follows:

$$I_{sp} = \frac{F}{\dot{m} g_0} \quad (26)$$

where $g_0$ is the standard gravity.

Fig. 14 describes the propulsion performance of the gas-solid two-phase rotating detonation engine with different particle diameters and hydrogen equivalence ratios. Fig. 14 (a) displays the thrust per unit area ($F$), with the results showing that $F$ increases as the hydrogen equivalence ratio decreases. Fig. 14 (b) shows the specific impulse ($I_{sp}$) with the results showing that $I_{sp}$ increases as the particle diameter decreases. The decrease in particle diameter leads to an increase in the reaction rate of the solid fuel and more participation of solid fuel in detonation combustion, resulting in higher thrust obtained per unit mass of propellant used.

In addition, in the specific impulse ($I_{sp}$) data, three points were observed to deviate from the general trend, as indicated by the arrows in the graph. When the particle diameter is 1 μm and the hydrogen equivalence ratio is 0.5, the specific impulse is 140.0 s, which is significantly lower than the cases with hydrogen equivalence ratios of 0.1-0.4. This is attributed to two reasons: Firstly, for this condition, the normalized wave height amplitude of the detonation wave is 0.544, significantly higher than other conditions, indicating a stronger instability of the detonation wave. Secondly, the case under this condition exhibits a two-phase double-front structure, while the other cases show a two-phase single-front structure, resulting in a lower intensity of the detonation wave for this particular case. The weaker and more unstable detonation wave leads to a reduction in specific impulse, resulting in a lower ability to generate thrust.

Similarly, when the particle diameter is 4 μm and 8 μm, the specific impulse of the cases with a hydrogen equivalence ratio of 0.3 is 131.3 s and 119.2 s, respectively, which are significantly lower than the cases with hydrogen equivalence ratios of 0.4-0.5. This is because, for larger particle diameters, the detonation wave is primarily sustained by hydrogen combustion due to the lower reaction rate of solid fuel. As a result, the main source of thrust is hydrogen, and a decrease in the hydrogen equivalence ratio leads to a noticeable decline in propulsion performance.

According to the research in this paper, particle diameter and hydrogen equivalence ratio are crucial physical parameters that significantly influence the detonation wave mode, propagation characteristics, stability, pressure gain, and propulsion performance in the gas-solid two-phase rotating detonation engine. Reducing the particle diameter enhances the speed and intensity of the two-phase detonation wave, improves the pressure gain in the combustion chamber, and increases the specific impulse. Decreasing the hydrogen equivalence ratio reduces the detonation wave speed, enhances the stability of the detonation flow field, increases the pressure gain in the detonation wave and combustion chamber, boosts thrust, and, in cases where the detonation structure is a two-phase single/double front, increases the intensity of the detonation wave or solid front.

To achieve optimal pressure gain and propulsion performance, it is advisable to minimize the particle diameter of solid fuel and reduce the hydrogen equivalence ratio while ensuring that the detonation wave exhibits a two-phase single-front structure. However, excessively small particle diameters may lead to a decrease in the stability of the detonation flow field. Therefore, in order to take into account the requirements of stability, pressure gain performance and propulsion performance, two-phase single-front detonation should be realized in gas-solid two-phase RDE, and smaller hydrogen equivalent ratio and appropriate particle diameter should be selected. According to the conclusion of this paper, the particle diameter should be 0.5-1 μm. Under such conditions, the detonation flow field demonstrates good stability, allowing the RDE to achieve higher pressure gain and specific impulse while maintaining stable operation. In practical applications, careful selection of the particle diameter of solid fuel should be made based on injection conditions, operating environment, and other relevant factors.

## 4. Conclusion

The focus of this paper is on the gas-solid two-phase rotating detonation engine. The flow field characteristics, wave structure, and instability of gas-solid two-phase RDE were investigated. The effects of solid fuel particle diameter and hydrogen equivalence ratio on the flow field characteristics and performance were revealed. The selection of operational conditions to ensure stable operation and optimal performance of the RDE was discussed. The main conclusions are summarized as follows:

1. Three flow field structures are identified in the gas-solid two-phase RDE. In the gas-solid double-front

structure, the gas-phase detonation wave propagates ahead, while the solid particles combust behind the detonation wave, forming a secondary high-pressure region. In the gas-solid single-front structure, the gas-phase detonation wave connects to the high-pressure region of the solid particles behind it. In the gas-phase single-front structure, the solid particles behind the gas-phase detonation wave combust slowly.

2. Reducing the hydrogen equivalence ratio and particle diameter both contribute to the transition from gas-phase single-front detonation to gas-solid two-phase double-front detonation and further to gas-solid two-phase single-front detonation. When the hydrogen equivalence ratio is too small and the solid fuel particle diameter is too large, the detonation wave cannot be self-sustained.

3. When the detonation wave exhibits a two-phase single/double-front structure, the detonation wave speed increases with the decrease in particle diameter. In most cases, the detonation wave speed decreases with the decrease in the hydrogen equivalence ratio. The detonation instability decreases with the decrease in the hydrogen equivalence ratio. When the hydrogen equivalence ratio is in the range of 0.1-0.2, the detonation instability increases with the decrease in particle diameter.

4. Reducing the particle diameter improves the pressure gain in the combustion chamber and increases the specific impulse. Decreasing the hydrogen equivalence ratio increases the pressure gain in the detonation wave and combustion chamber and boosts thrust. To achieve the best pressure gain and propulsion performance, efforts should be made to minimize the particle diameter and hydrogen equivalence ratio of the solid fuel while ensuring a two-phase single-front detonation wave. However, excessively small particle diameter may lead to decreased stability of the detonation flow field.

5. When selecting working conditions for gas-solid two-phase rotating detonation engines, stability and propulsion performance requirements should be considered. The value of the solid-phase fuel particle diameter, $d$, should be carefully chosen, provided that two-phase single-front detonation is achieved and a small hydrogen equivalence ratio is selected. Lowering the value of $d$ can enhance the pressure gain and specific impulse of the RDE but may reduce the stability of the rotating detonation flow field, which is not favorable for stable engine operation. The value of $d$ should be chosen to ensure stable propagation of the rotating detonation wave while maximizing the pressure gain in the combustion chamber and specific impulse of the engine.

## Acknowledgments

This research is sponsored by the National Natural Science Foundation of China (Grant No. 91741202 and No. 52076003). We are also very grateful to the excellent editors and reviewers for their guidance and suggestions on this manuscript, which have improved the clarity, scientific rigor, and accuracy of this manuscript.

## Data Availability

The data that support the findings of this study are available from the corresponding author upon reasonable request.

## Appendix

**1. More quantitative analysis of detonation mode transition process**

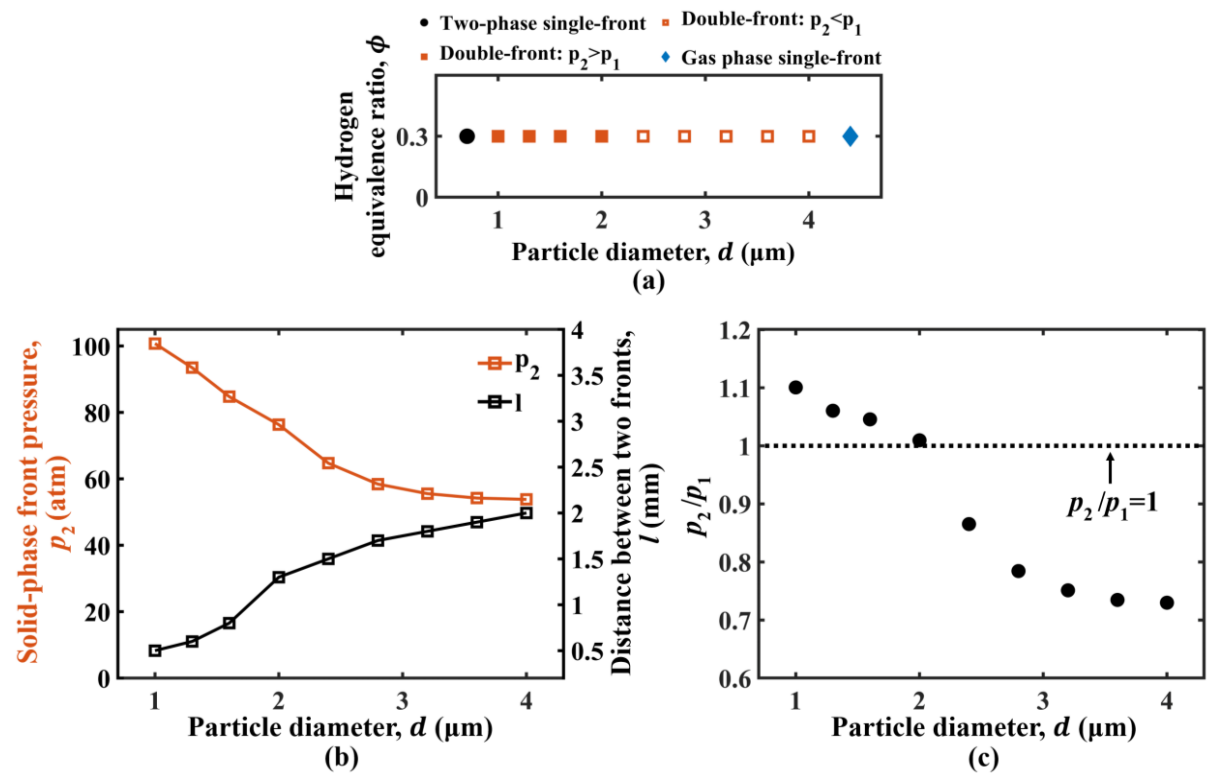

**Fig. A1 Quantitative analysis of the transition process between the gas-solid two-phase rotating detonation modes. (a) Case operating modes. (b) The pressure at the solid-phase front, $p_2$, and the distance between the two fronts, $l$, are plotted against particle diameter. (c) The ratio of the pressure at the solid-phase front to the pressure at the detonation front, $p_2/p_1$, is shown as a function of particle diameter.**

To further clarify the transition process from the two-phase single-front structure to the two-phase double-front structure and finally to the gas-phase single-front structure, quantitative analysis was conducted for 11 cases with particle diameters ranging from 0.7 μm to 4.4 μm at a hydrogen equivalence ratio of 0.3. Fig. A1(a) shows the operating conditions for these 11 cases. The two-phase single-front structure is represented by black solid circles, cases where the solid-front pressure is greater than the detonation front pressure ($p_2/p_1>1$) are represented by red solid squares, cases where the solid-front pressure is less than the detonation front pressure ($p_2/p_1<1$) are represented by red hollow squares, and the gas-phase single-front structure is represented by blue solid diamonds.

Figs. A1(b)-(c) show the variations of the solid-front pressure $p_2$, distance between the two fronts $l$, and the ratio of solid-front pressure to detonation front pressure $p_2/p_1$ with particle diameter. According to the analysis in Section 3.1, the two-phase single-front structure is essentially a result of the connection or coincidence between the gas-phase front and the solid-phase front due to the higher solid reaction rate. As the particle diameter increases, the solid-phase high-pressure region gradually separates from the detonation front, and an independent solid-phase front forms, leading to the transition from the two-phase single-front structure to the two-phase double-front structure.

When the particle diameter is greater than 1 μm, with further increase in diameter, the solid-front pressure $p_2$ gradually decreases, the distance between the two fronts $l$ increases, and $p_2/p_1$ decreases. When the particle diameter exceeds 4 μm, the solid-phase high-pressure region disappears, and the detonation front takes on a gas-phase single-front structure.

## 2. Quantitative analysis of the formation of counter-rotating shock waves

Next, a quantitative analysis of the formation of counter-rotating shock waves will be conducted. The interface between fresh gas and combustion products exhibits a serrated pattern, with fresh gas and combustion products alternating in arrangement. At these serrated boundaries, the rotating detonation wave will alternately pass through fresh gas and combustion products during propagation.

The temperature of the fresh gas at the boundary is about 370 K and the density is about 3.8 kg/m$^3$, and the temperature of the combustion product is about 2330 K and the density is about 0.6 kg/m$^3$. The reactant has a low temperature and high density, while the combustion products have a high temperature and low density. The pressure of the reactant and products is approximately equal.

As the rotating detonation wave propagates near the serrated interface, it continuously collides with the reactant and product contact surfaces in the local region. In these collision processes, a series of small-scale counter-rotating shock waves are generated.

The formation of small-scale counter-rotating shock waves can be explained by the reflection principle of shock waves passing through different medium interfaces [72]. Let the shock wave (detonation wave) propagate from left to right, and consider two media separated by an interface: the left medium with a specific heat ratio of $\gamma_1$ and specific volume of $v_1$, and the right medium with a specific heat ratio of $\gamma_2$ and specific volume of $v_2$.

When $\gamma_1/v_1 < \gamma_2/v_2$ and $(\gamma_1+1)/v_1 < (\gamma_2+1)/v_2$, the shock wave colliding with the interface will produce a reflected shock wave propagating to the left. Conversely, when $\gamma_1/v_1 > \gamma_2/v_2$ and $(\gamma_1+1)/v_1 > (\gamma_2+1)/v_2$, the reflected wave will be an expansion wave propagating to the left.

In the case of the rotating detonation wave passing through the serrated fresh gas interface, there are two interactions. The first one is the detonation wave propagating from the high-temperature and low-density products to the low-temperature and high-density reactants. The second one is the detonation wave propagating from the low-temperature and high-density reactants to the high-temperature and low-density products. The specific heat ratios of the reactants and products near the serrated interface are 1.39 and 1.24, respectively, indicating that their specific heat ratios are close. The pressure of the products and reactants is approximately equal, and the ratio of specific volumes is about 6.3.

When the detonation wave propagates from the products to the reactants, where the subscripts 1 and 2 correspond to the products and reactants, respectively, the specific volume ratio is $v_1/v_2 \approx 6.3$. By substituting the data into the inequality, it is evident that $\gamma_1/v_1 < \gamma_2/v_2$ and $(\gamma_1+1)/v_1 < (\gamma_2+1)/v_2$ always hold true, indicating that the reflected wave in this case is a shock wave propagating to the left.

On the other hand, when the detonation wave enters the products from the reactants, where the subscripts 1 and 2 correspond to the reactants and products, respectively, the specific volume ratio is $v_1/v_2 \approx 1/6.3$. By substituting the data into the inequality, it is evident that $\gamma_1/v_1 > \gamma_2/v_2$ and $(\gamma_1+1)/v_1 > (\gamma_2+1)/v_2$ always hold true, indicating that the reflected wave in this case is an expansion wave propagating to the left.

In summary, as the rotating detonation wave propagates near the serrated fresh gas interface and continuously transitions from the products to the fresh gas, small-scale counter-rotating shock waves are repeatedly generated in the opposite direction to the detonation wave propagation. These counter-rotating shock waves can affect the intake process and enhance the instability of the detonation wave and the combustion chamber.